\font\manual=manfnt
\def\dbend{{\manual\char127}}

\def\res{\mathop{\rm res}\nolimits}
\def\disc{\mathop{\rm disc}\nolimits}
\def\lc{\mathop{\rm lc}\nolimits}

\def\RootOf{\mathop{\rm RootOf}\nolimits}

\def\arccosh{\mathop{\rm arccosh}\nolimits}

\def\A{{\mathbf A}}

\def\LL{{\cal L}}
\def\LRCF{{\LL_{\rm RCF}}}
\def\C{{\mathbf C}}

\def\PP{{\cal P}}
\def\Q{{\mathbf Q}}
\def\R{{\mathbf R}}

\def\ii{{\bf i}}
\def\jj{{\bf j}}
\def\eqq{{\buildrel?\over=}}

\documentclass[runningheads,a4paper]{llncs}
\usepackage{amssymb}
\usepackage{graphicx}
\usepackage{url}
\usepackage[show]{ed}

\usepackage{algorithm2e}
\usepackage{qtree}

\begin{document}

\mainmatter


\title{Recent Advances in Real Geometric Reasoning}  
\titlerunning{Recent advances in real geometric reasoning}
\author{James H. Davenport$^1$ \and Matthew England$^2$}
\authorrunning{James H.~Davenport and Matthew England}
\institute{$^1$Departments of Computer Science \& Mathematical Sciences,
University of Bath, UK \\
$^2$ Department of Computing, Coventry University, UK \\
\email{J.H.Davenport@bath.ac.uk, Matthew.England@coventry.ac.uk}
}

\maketitle
\begin{abstract}
In the 1930s Tarski showed that real quantifier elimination was possible, and in 1975 Collins gave a remotely practicable method, albeit with doubly-exponential complexity, which was later shown to be inherent. We discuss some of the recent major advances in Collins method: such as an alternative approach based on passing via the complexes, and advances which come closer to ``solving the question asked'' rather than ``solving all problems to do with these polynomials''.
\end{abstract}

\section{Introduction}

Although methods with better asymptotic complexity are known in theory (e.g. \cite{GrigorievVorobjov1988a}), the workhorse of implemented algorithms for real geometric reasoning is Cylindrical Algebraic Decomposition. This was introduced in \cite{Collins1975} to produce a remotely practicable (complexity ``merely'' doubly exponential in the number of variables) alternative to Tarski's original method from 1930 \cite{Tarski1951}, whose complexity could not be bounded by any tower of exponentials. Tarski in fact set out to solve the \emph{quantifier elimination problem} for real algebraic geometry (Section \ref{sec:QE}): given $Q_{k+1} x_{k+1}Q_{k+2} x_{k+2}\dots \Phi(x_1,\ldots,x_n)$, where $Q_i\in\{\forall,\exists\}$ and $\Phi$ is a Boolean combination of relations involving polynomials $p_i(x_1,\ldots,x_n)$, find an equivalent $\Psi(x_1,\ldots,x_k)$, where $\Psi$ is a Boolean combination of relations involving polynomials $q_i(x_1,\ldots,x_k)$. In fact, we cannot solve this in the language of algebraic geometry: we need \emph{semi-algebraic} geometry, allowing $>$ as well\footnote{Strictly speaking $>$ is sufficient, but implementations always allow $\ge$ and $\ne$. In fact, $\ne$ is intrinsic to the regular chains approach discussed in Section \ref{sec:RC}.} as $=$. The \emph{necessity} of $>$ follows from the fact of the example $\exists y: x=y^2  \Leftrightarrow (x>0)\lor(x=0)$; its \emph{sufficiency} is the point of Tarski's work.

\section{Cylindrical Algebraic Decomposition by Projection and Lifting}
\label{sec:PL}

\cite{Collins1975} constructs a sampled\footnote{The ``sampled'' nature is implicit in \cite[and its successors]{Collins1975}, but the authors find it helpful to be explicit about this.} \emph{Cylindrical Algebraic Decomposition} (CAD) of $\R^n$ which is \emph{sign-invariant} for the $p_i$, where these words are defined as follows.
\begin{definition}[CAD terminology]\label{def:cad}
Note that throughout we are ordering our coordinates/\allowbreak{}vari\-ables, so that $x_n$ is the ``last coordinate''.
\begin{description}
\item[decomposition:] a partition of $\R^n$ into cells $C_\ii$ indexed by $n$-tuples of natural numbers (so $\R^n=\bigcup_\ii C_\ii$ and $\ii\ne\jj\Rightarrow C_\ii\cap C_\jj=\emptyset)$;
\item[{\rm(semi-)}algebraic:] every $C_\ii$ is defined by a finite set of equalities and inequalities of polynomials in the $x_i$, including expressions of the form
\begin{equation}\label{eq:RootOf}
\RootOf_2(f_1(x_1,y))<x_2<\RootOf_3(f_2(x_1,y)) 
\end{equation}
(where $\RootOf_2$ means ``the second real root, counting from $-\infty$'');
\item[cylindrical:] for all $k<n$, if $\pi_k$ is the projection onto the \emph{first} $k$ coordinates, then, for all $\ii,\jj$, $\pi_k(C_\ii)$ and $\pi_k(C_\jj)$ are either equal or disjoint;
\item[sampled:] for each cell $C_\ii$ there is an explicit point $s_\ii\in C_\ii$;
\item[sign-invariant:] for the polynomials in $\Phi$ on each cell, every $p_i$ is identically zero, or everywhere positive, or everywhere negative.
\end{description}
\end{definition}
Collins constructed such a decomposition by a process now known (at least by our colleagues) as \emph{CAD by Projection and Lifting} (for more details see \cite{Davenport2015c}).  The key property in this approach is the following.
\begin{definition}\label{def:del}
A polynomial $p(x_1,\ldots,x_m)$ is \emph{delineable}\footnote{There are various, subtly different, definitions in the literature. This one is from \cite{McCallum1999a}.} over a region $C\subset\R^{m-1}$ if:
\begin{enumerate}
\item the portion of the real variety of $p$ that lies in the cylinder $C\times\R$ over $C$ consists of the union of the graphs (called sections) of some $k\ge0$ continuous functions $\theta_1<\cdots<\theta_k$ from $C$ to $\R$ {\bf and};
\item there exist integers $m_1,\ldots, m_k \ge 1$ such that, for every point $(a_1,\ldots, a_{m-1})$ in $C$, the multiplicity of the root $\theta_i(a_1,\ldots, a_{m-1})$ of $p(a_1,\ldots, a_{m-1},x_m)$, considered as a function of $x_m$ alone, is $m_i$ (and in particular is constant).
\end{enumerate}
A set of polynomials is delineable over $C$ if each is delineable {\bf and} if the sections are either identical or disjoint. This is actually equivalent to saying that the product is delineable.
\end{definition}
Intuitively, if the $\{p_i\}$ are delineable over $C$, their graphs neither fold nor cross.
\par
Let $\PP_n$ be the set of polynomials in $\Phi$, with coefficients from some effective\footnote{The literature often stipulates $\Q$ or the algebraic numbers $\A$. The real requirement is that we can perform all the polynomial algebra we need over $K$, and that, given expressions $a,b\in K$, we can decide the trichotomy $a<b$ or $a=b$ or $a>b$. Once we start adding transcendental functions to our language, the effectivity of $K$ becomes a major problem, as we run across the usual indecidability results. This is addressed in different ways in \cite{Achatzetal2008} and \cite{Vorobjov1989,Vorobjov1992}.} field $K\subset\R$.  Then Collins algorithm proceeds as follows:
\begin{description}
\item[Projection:] Given some $\PP_k\subset K[x_1,\ldots,x_k]$ construct a set $\PP_{k-1}\subset K[x_1,\ldots,\allowbreak{}x_{k-1}]$ such that, over each cell of a CAD sign-invariant for $\PP_{k-1}$, the polynomials of $\PP_k$ are delineable. Though the details depend on the algorithm, the key ingredients are leading coefficients (where these vanish some $\theta_i$ tends to infinity), discriminants (where these vanish some $\theta_i$ ceases to have constant multiplicity)  and resultants (where these vanish, the $\theta_i$  from different polynomials intersect).  
\\
Repeat until we have the set of univariate polynomials $\PP_1$.
\item[Base case:] Given $\PP_1$, isolate the $N_1$ real roots of these polynomials in $\R^1$, and construct a CAD consisting of the $N_1$ roots, and the $N_1+1$ intervals between them (or to the left/right of them all). The sample points for the 0-dimensional cells are the roots themselves: for the 1-dimensional intervals we choose any convenient point, generally rational and with denominator the smallest power of 2 we can find.
\item[Lifting:] Given a CAD $D_{k-1}$ of $\R^{k-1}$, sign-invariant for $\PP_{k-1}$, construct a CAD $D_{k}$ of $\R^{k}$, sign-invariant for $\PP_{k}$. For each cell $C_\ii$, this is done by substituting the sample point $s_\ii$ into $\PP_k$, and doing the equivalent of the base case for the resulting univariate system (valid across the whole of $C_\ii$ if the projection operator provides delineable projection polynomials).
\\
Repeat until we have the CAD $D_{n}$ of $\R^{n}$.
\end{description}
If we suppose that $\PP_n$ contains $m$ polynomials, of degree (in each variable) bounded by $d$, and coefficient length bounded by $l$ (coefficients bounded by $2^l$), then the time  complexity is bounded  \cite[Theorem 16]{Collins1975} by
\begin{equation}\label{eq:Collins}
O\left(m^{2^{n+6}}(2d)^{2^{2n+8}}l^3\right).
\end{equation}\def\foo#1#2{\hbox to 0pt{/\hskip-6pt}#1{#2}}%
This analysis is very sensitive to the details of the sub-algorithms involved, and a more refined analysis of the base case \cite{Davenport1985c} reduces the complexity (though not the actual running time) to 
\[
O\left( m^{2^{n+\foo64}} (2d)^{2^{2n+\foo86}} l^3 \right).
\]
This improvement might seem trivial, but in fact implies taking the fourth root of the $m,d$ part of the complexity. 
\par
A less sensitive property (and one that reflects the cost of \emph{using} such a decomposition) is the number of cells: for the Collins method this is bounded, by an analysis similar to \cite{Bradfordetal2014a}, by
\begin{equation}\label{eq:Collinscell}
O\left(m^{2^{n}}(2d)^{2\cdot3^{n}}\right).
\end{equation}
As is often the case in mathematics, we get more insight if we solve an apparently harder problem. \cite{McCallum1984} did this, demanding that the decompositions $D_k$, $k<n$ be, not just sign-invariant, but 
\begin{description}
\item[order-invariant] for the polynomials in $\Phi$, i.e. on each cell, every $p_i$ is identically zero, and vanishes to the same order throughout the cell, or everywhere positive, or everywhere negative.
\end{description}
This actually lets his $\PP_k$ be much simpler than Collins', with the cost that the lifting procedure might fail if some element $p$ of $\PP_k$ \emph{nullifies} (is identically zero) over some cell in $D_{k-1}$.  In this case, McCallum says that $\PP_k$ was not \emph{well-oriented}, and has to either:
\begin{enumerate}
\item proceed by working around the problem or concluding it not relevant.  This is only possible in certain cases (e.g. the cell is dimension 0) \cite{Brown2005}.  Otherwise;
\item revert to Collins' projection (or a variant due to \cite{Hong1990b}); or,
\item add the partial derivatives of $p$ to $\PP_k$ and resume the projection process from there --- an operation that to the best of the authors' knowledge has never been implemented, doubtless because of the complicated backtracking involved, and the fact that, whereas we only \emph{ought} to add this polynomial in the nullifying region, the design of Collins' algorithm and its successors assume a global set of polynomials at each level.
\end{enumerate}
``Randomly'', well-orientedness ought to occur with probability 1,
but we have a family of ``real-world'' examples (simplification/branch cuts, see \cite{Beaumontetal2007}) where it often fails.
The analogy of (\ref{eq:Collinscell}) is given by \cite[Theorem 6.1.5]{McCallum1985b} as 
\begin{equation}\label{eq:McCcell}
O\left(m^{2^{n}}(2d)^{n\cdot2^{n}}\right),
\end{equation}
and a recent improved analysis in \cite[(12)]{Bradfordetal2014a} reduces this to
\begin{equation}\label{eq:McCcell-us}
O\left(2^{2^{n-1}}m(m+1)^{2^{n}-2}d^{2^{n}-1}\right).
\end{equation}

\section{CAD by Regular Chains}
\label{sec:RC}

This alternative to the traditional computation scheme of projection and lifting was introduced in \cite{Chenetal2009d}, then improved in \cite{ChenMorenoMaza2012a}.  The method can be described as ``going via the complexes'', since the authors first construct a cylindrical decomposition of $\C^n$, and then infer a CAD of $\R^n$.  They make use of the well developed body of theory around regular systems \cite{Wang2000} for the work over the complexes, and the algorithms are all implemented in the \texttt{RegularChains} Library\footnote{\texttt{http://www.regularchains.org}} for \textsc{Maple}, hence our designation: \emph{CAD by Regular Chains}.

We first need the following analogue of Definition \ref{def:del} (not precisely analogous, as Definition \ref{def:del} allows for non-square-free polynomials and this does not).
\begin{definition}\label{def:sep}Let $K\subset\C$ be an effective field.
Let $C$ be a subset of $\C^{n-1}$ and $P \subset  K[x_1,\ldots,x_{n-1}, x_n]$ be a finite set of polynomials whose main variable really is $x_n$. We say that $P$ \emph{separates above} $C$ if for each $\alpha\in C$:
\begin{enumerate}
\item for each $p\in P$, the polynomial $\lc_{x_n}(p)$ does not vanish at $\alpha$;
\item the polynomials $p(\alpha, x_n) \in \C[x_n]$, for all $p\in P$, are squarefree and coprime.
\end{enumerate}
Note that the empty set is trivially separable.
\end{definition}
\def\cD{{\cal D}}
We then need an analogue of Definition \ref{def:cad} for the case of complex space.  We follow \cite{ChenMorenoMaza2012a} and describe these (complex) cylindrical decompositions in terms of the tree data structure they are stored as.
\begin{definition}\label{def:ccd}
We define a \emph{cylindrical decomposition} of $\C^n$, and its associated tree, by induction on $n$.
\begin{description}
\item[Base Either:] There is one set $D_1$, the whole of $\C$ and $\cD=\{D_1\}$;
\item[Base Or:] there are $r$ non-constant square-free relatively prime polynomials $p_i$ such that $D_i$ is the set of zeros of $p_i$, and $D_{r+1}$ is the complement: $\{x: p_1(x)p_2(x)\cdots p_r(x)\ne0\}$: $\cD=\{D_1,\ldots,D_r,D_{r+1}\}$.
\item[Base Tree:] The root and all the $D_i$ as leaves of it.
\item[Induction:] Let $\cD'$ be a cylindrical decomposition of $\C^{n-1}$. For each $D_i\in\cD'$, let $r_i$ be a non-negative integer, and $P_i=\{p_{i,1},\ldots,p_{i,r_i}\}$ be a set of polynomials which separates over $D_i$. 
\item[Induction Either:] $r=0$ and we set $D_{i,1}=D_i\times\C$;
\item[Induction Or:] we set $D_{i,j}=\{(\alpha,x): \alpha\in D_i \land p_{i,j}(\alpha,x)=0\}$; 
\item[\qquad\quad]$D_{i,r+1}=\left\{(\alpha,x): \alpha\in D_i \land \prod_jp_{i,j}(\alpha,x)\ne0\right\}$;
\item[Then:] a cylindrical decomposition of $\C^n$ is given by
\[
\cD=\{D_{i,j}: 1\le i\le |\cD'|; 1\le j\le r_i+1\}.
\]
\item[Induction Tree:] If $T'$ is the tree associated to $\cD'$ then the tree associated to $\cD$ is obtained by adding to each leaf $D_i \in T'$ as children all the $D_{i,j}$ such that $1\le j\le r_i+1$.
\end{description}
\rm Unlike Definition \ref{def:cad}, the different roots of a given polynomial are not separated. Each cell is the zero set of a system of polynomial equations and inequations, where the main variables are all distinct: a triangular system \cite{Aubryetal1999}.
\end{definition}
\begin{definition}\label{def:ccd-f}Let $F$ be a set of polynomials in $k=K[x_1,\ldots,x_n]$. A cylindrical decomposition $\cD$ is \emph{$F$-invariant} if, for each cell $D\in\cD$ and each $f_i\in F$, either $f$ vanishes at all points of $D$ or $f$ vanishes at no point of $D$.
\end{definition}
The trivial decomposition, obtained by taking the ``either'' branch each time, with one cell, is $\emptyset$-invariant. Given a cylindrical decomposition $\cD$ which is $F$-invariant, and supposing $\widehat F=F\cup \{f\}$, \cite{ChenMorenoMaza2012a} shows how to refine $\cD$ to a cylindrical decomposition $\widehat\cD$ which is $\widehat F$-invariant, hence the ``incremental'' in the title of their paper. The key ingredients in this process are again leading coefficients, resultants and discriminants. 
The paper \cite{Chenetal2009d} shows, assuming that $K\subset\R$, how to construct from $\cD$ a cylindrical algebraic decomposition of $\R^n$ which is sign-invariant for $F$.
\par
The construction of the cylindrical decomposition can be seen, as pointed out in \cite{ChenMorenoMaza2012a}, as an analogue of the projection phase of projection and lifting. Indeed, if $n$ is small, it is often the case that the polynomials at level $i$ in the tree corresponding to $\cD$ are those in $\PP_{n-i}$. The fundamental difference is that the $\PP_i$ are \emph{global} structures: over the whole cylindrical algebraic decomposition of $\R^k$ we need to isolate all the branches of all of $\PP_{k+1}$, whereas there is a tree structure underpinning $\cD$ and the cylindrical algebraic decomposition, which means that ``polynomials are not considered when they are blatantly not relevant''.

\newpage

\noindent \textbf{Example:} Consider the parabola $p := ax^2 + bx + c$ and assume the variable ordering $x \succ c \succ b \succ a$.  Suppose we were to use projection and lifting.  Then the first projection identifies the coefficients $a,b,c$ and the discriminant with respect to $x$: $b^2-4ac$.  Subsequent projection do not identify any further projection polynomials for this example.  Lifting produces CADs sign-invariant for these 4 projection polynomials, as well as $p$ itself.  

The regular chains approach would start by building the following tree, representing a cylindrical decomposition of $\C^n$:

\begin{center}
\qtreecenterfalse
\Tree [. [.{$a=0$} [.{$b=0$} {\quad$c=0$} {\quad$c\neq0$} ] [.{$b\neq0$} {\quad$p=0$} {\quad$p\neq0$} ] ] 
      [.{$a\neq0$} [.{$b^2-4ac=0$} {\quad$p=0$} {\quad$p\neq0$} ] [.{$b^2-4ac\neq0$} {\quad$p=0$} {\quad$p\neq0$} ] ] ]
\end{center}

This decomposition was produced to be sign-invariant for $p$.  However, it does not insist on sign-invariance for the all the other projection polynomials.  In particular, it is not sign-invariant for $b$.  The polynomial $b$ is included in the projection set because its vanishing can determine delineability, but only when the coefficient of the higher degree terms vanish.  So, when $a=0$ it is important to ensure $b$ is sign-invariant, but not otherwise.  Hence the tree above is doing only what is necessary to make the final conclusion about $p$.  

The next step is to apply real root isolation, extending this tree to one representing a CAD.  At the top level the case $a \neq 0$ must split into the two possibilities: $a<0$ and $a>0$.  For brevity we display only the branch for $a<0$ below (where $r_1$ and $r_2$ represent the two real roots of $p$ in the case where the leading coefficient is negative and the discriminant positive).  The full tree has 27 leaves, thus representing a CAD with 27 cells.  This compares with a minimal CAD of 115 cells produced by projection and lifting to be sign invariant for all projection polynomials.


\begin{center}
\qtreecenterfalse
\treewidth=3cm
\Tree [. 
[.{$a<0$} 
	[.{$c=\frac{b^2}{4a}$} {$x<-\frac{b}{2a}$} {$x=-\frac{b}{2a}$} {$x>-\frac{b}{2a}$} ] 
	[.{\vline} [.{$c>\frac{b^2}{4a}$} {$x<r_1$} {$x=r_1$} {$x \in (r_1,r_2)$} {$x=r_2$} {$x>r_2$} ] ]
	!\qsetw{5cm}
	[.{$c<\frac{b^2}{4a}$} ] 
]
]
\end{center}

\par
Are there significant savings in general? We refer the reader to \cite[Table 1]{Bradfordetal2014a}.  Here {\tt PL-CAD} refers to our implementation of McCallum's algorithm of Section \ref{sec:PL}; {\tt RC-INC-CAD} refers to the algorithm of \cite{ChenMorenoMaza2012a} (Section \ref{sec:RC}); and \textsc{Qepcad} \cite{Brown2003} is another, highly optimised, implementation of McCallum's algorithm. Where both terminate, \textsc{Qepcad} and {\tt PL-CAD} often, though not always, have the same cell count. {\tt RC-INC-CAD} does sometimes have the same count, but on other examples such as BC-Phisanbut-4, needs only 2007 cells, while both implementations of McCallum's algorithm need 51,763.

\section{Quantifier Elimination}
\label{sec:QE}

The original motivation for \cite{Collins1975} was the following problem.
\begin{problem}[Quantifier Elimination]Let $Q_i\in\{\exists,\forall\}$, and $\LRCF$ be the language of Boolean-connected equalities and inequalities concerning polynomials in $K[x_1,\allowbreak\ldots,x_n]$, where $K$ is an effective field with $\Q\subseteq K\subset\R$ . Given a statement (known as a Tarski statement, or a Tarski sentence if $k=0$)
\begin{equation}\label{eq:Q}
\Phi:=Q_{k+1}x_{k+1}\ldots Q_nx_n \phi(x_1,\ldots,x_n): \qquad \phi\in \LRCF,
\end{equation}
the \emph{Quantifier Elimination} problem is that of producing an equivalent
\begin{equation}\label{eq:QF}
\Psi:=\psi(x_1,\ldots,x_k) : \qquad \psi\in \LRCF.
\end{equation}
In particular, $k=0$ is a decision problem: is $\Phi$ true?
\end{problem}
If we have a CAD $\cD^{(n)}$ of $\R^n$ (noting that the $x_i$  must be ordered in the same way in Definition \ref{def:cad} and formula (\ref{eq:Q})) sign-invariant for the polynomials of $\Phi$, then constructing $\Psi$ is conceptually easy.
\begin{enumerate}
\item The truth of $\phi$ in a cell $D_\ii$ of $\cD^{(n)}$  is that of $\phi$ at the sample point $s_\ii$.\label{first}
\item $\cD^{(n)}$ projects to a  CAD $\cD^{(k)}$ of $\R^k$.
\item The truth of $\Phi$ in a cell $\widehat D_\jj\in \cD^{(k)}$ is then the appropriate ($\bigvee$ for $\exists$ etc.) Boolean combination of the truth of $\phi$ in the cells of $\cD$ that project to $\widehat D_\jj$.\label{step:bool}
\item $\Psi$ is then the disjunction of the defining formulae for all the $\widehat D_\jj$ for which $\Phi$ is true.\label{last}
\end{enumerate}
There is a problem in practice with the last step, first pointed out in \cite{Brown1999a}. In the lifting stage, we produce branches $\theta_i$ of polynomials, with descriptions such as ``that branch of $p(x_1,\ldots,x_l)$ which, above the sample point $s=(\alpha_1,\ldots,\alpha_{l-1})$, has the (unique) root in $(\beta,\gamma)$'', and this is not a statement of $\LRCF$. We could equally describe it as ``the third real branch of $p(x_1,\ldots,x_l)$ above $s$'', but again this statement is not in $\LRCF$. Now by Thom's Lemma \cite{CosteRoy1988}, we can describe this branch in terms of the signs of $p$ \emph{and its derivatives}, but, whereas these derivatives are in the Collins projection, they are not in the McCallum projection, or in the tree constructed by the method of Section \ref{sec:RC}. However, when it comes to describing $\widehat D_\jj$, we can just add these (as described in \cite{Brown1999a} for projection and lifting and in \cite{CM14} for regular chains CAD construction). The additional cost is negligible, in particular, we do not need them for projection (Section \ref{sec:PL}), or for tree construction (Section \ref{sec:RC}).

Though it may depend non-linearly on polynomial degree etc., this process is linear in the number of cells in $\cD^{(n)}$, and produces a disjunction of at most as many clauses as there are cells in $\cD^{(k)}$.

\section{Lower Bounds}
\label{sec:LB}

This last remark is the basis of the complexity lower bounds in \cite{DavenportHeintz1988,BrownDavenport2007}. Both constructions use the fact that
\begin{equation}\label{eq:DH}
\exists z_m\forall x_{m-1}\forall y_{m-1}
\vtop{\vskip-2.0ex$\hskip-10pt\left( \begin{array}{c}
\left(y_{m-1}=y_m\land x_{m-1}=z_m\right)\lor\left(y_{m-1}=z_m\land x_{m-1}=x_m\right)\cr
\Rightarrow y_{m-1}=F_{m-1}(x_m{-1})\end{array}\right)$\hskip-100pt}
\end{equation}
encodes $y_m=F_{m-1}(F_{m-1}(x_m))$. Hence applying this construct $m-1$ times to $y_1=F_1(x_1)$ gives 
\[
y_m=\underbrace{F_1(F_1(\cdots F_1(}_{\hbox{$2^{m-1}$ times}}(x_n))\cdots).
\]
This can then be used to produce expressions with $n$ quantifiers and having $2^{2^{O(n)}}$ isolated point solutions, hence needing $2^{2^{O(n)}}$ cells to describe them (the $O(n)$ terms are $n/3+O(1)$ in \cite{BrownDavenport2007} and $n/5+O(1)$ in \cite{DavenportHeintz1988}). 
An example which needs  $2^{2^{O(n)}}$ cells for all possible variable orders is also 
produced in \cite{BrownDavenport2007}, along with another which needs $2^{2^{O(n)}}$ cells in one order, but a constant number in another.  Hence the great interest in variable order selection methods for CAD \cite[to name a few]{Dolzmannetal2004a,Englandetal2014c,Huangetal2014a}.
\par
The construction in (\ref{eq:DH}) uses both $\exists$ and $\forall$ in a way that cannot be unnested. In fact, it is possible \cite{Grigoriev1988} to decide Tarski sentences (i.e. no free variables) with a cost that is singly-exponential in $n$, but doubly-exponential in $a$, the number of \emph{alternations} of $\exists$ and $\forall$ in (\ref{eq:Q}). These methods, or any methods singly-exponential in $n$, have, in general, not been implemented, though there has been work on the purely existential case (for example \cite{Huntington2008a}).

\section{Equational Constraints}
\label{sec:EC}

The methods described in the previous sections produce decompositions which are sign- (or order-)invariant for a set of polynomials.  In particular, we can apply steps \ref{first}--\ref{last} of Section \ref{sec:QE} to the same CAD to solve (\ref{eq:Q}) for any other $\phi$  involving the same polynomials. Indeed, as long as the $x_i$ stayed in the same order, we could change the $Q_i$ as well. \cite{Collins1998} suggested that we could do better if $\phi$ was of the form $p_1=0\land\phi'$, as we would not be interested in the behaviour of polynomials in $\phi'$ except when $p_1=0$. This was implemented in \cite{McCallum1999a}, who produced a CAD which was sign-invariant for $p_1$, and {\bf when $p_1=0$}, sign-invariant for the polynomials in $\phi'$. The main effect of this is to reduce the double exponent $n$ of $m$ in (\ref{eq:McCcell-us}) by 1, i.e. to take the square root of this term, as shown recently in 
\cite{Bradfordetal2014a} (14).
\par
It is worth seeing how this works. Consider
\begin{equation}\label{eq:EC}
\phi:=(f_1=0)\land\left((f_2>0)\lor(f_3>0)\right).
\end{equation}
Then a \cite{McCallum1984}-style projection ignoring the fact that there is an equation constraint would contain\footnote{It would also have some leading coefficients etc., but these are not the main drivers of the complexity in McCallum's projection.} three $\disc(f_i)$ and three $\res(f_i,f_j)$. However, \cite{McCallum1999a} observes that we are not interested in $f_2,f_3$ except when $f_1=0$, and hence we need only consider $\disc(f_1)$ and $\res(f_1,f_2)$, $\res(f_1,f_3)$, half as many polynomials.

\par Consider now
\begin{equation}\label{eq:TTI}
\phi_1:=\left((g_1=0)\land(g_2>0)\right)\lor\left((g_3>0)\land(g_4=0)\right).
\end{equation}
A \cite{McCallum1984}-style projection ignoring the fact that there is an equation constraint would contain four discriminants  and six resultants.
Although (\ref{eq:TTI}) does not contain an \emph{overt} equational constraint,  $\phi_1\Rightarrow(g_1=0)\lor(g_4=0)$, which is $\phi_1\Rightarrow(g_1g_4=0)$, and so the equational constraint $g_1g_4=0$ is implicit. If we study $g_1g_4=0\land\phi_1$ in the style of (\ref{eq:EC}), and drop trivial resultants, we consider $\disc(g_1g_4)$, $\res(g_1g_4,g_2)$ and  $\res(g_1g_4,g_3)$. Using the multiplicative properties of resultants and discriminants (which we would certainly do in practice!), this is $\disc(g_1)$, $\disc(g_4)$ and all the resultants except $\res(g_2,g_3)$, i.e. two discriminants and five resultants.
\par
Intuitively $\res(g_1,g_3)$ and $\res(g_2,g_4)$ are redundant, but how do we achieve this in general? This was solved in \cite{Bradfordetal2013b}, where, rather than producing a sign-invariant CAD, we compute  \emph{truth table invariant} (a TTICAD) for the two propositions $(g_1=0)\land(g_2>0)$ and $(g_3>0)\land(g_4=0)$, i.e. on each cell, each of these two propositions is either identically true, or identically false. This process does indeed remove these two resultants, so we have two discriminants and three resultants.

\vspace*{0.1in}

\begin{figure}[t]
\caption{The left is a sign-invariant CAD, and the right a TTICAD, for (\ref{eq:TTI}) with the polynomials from the example.}
\label{fig:Ex1}
\centering
\includegraphics[width=0.48\textwidth]{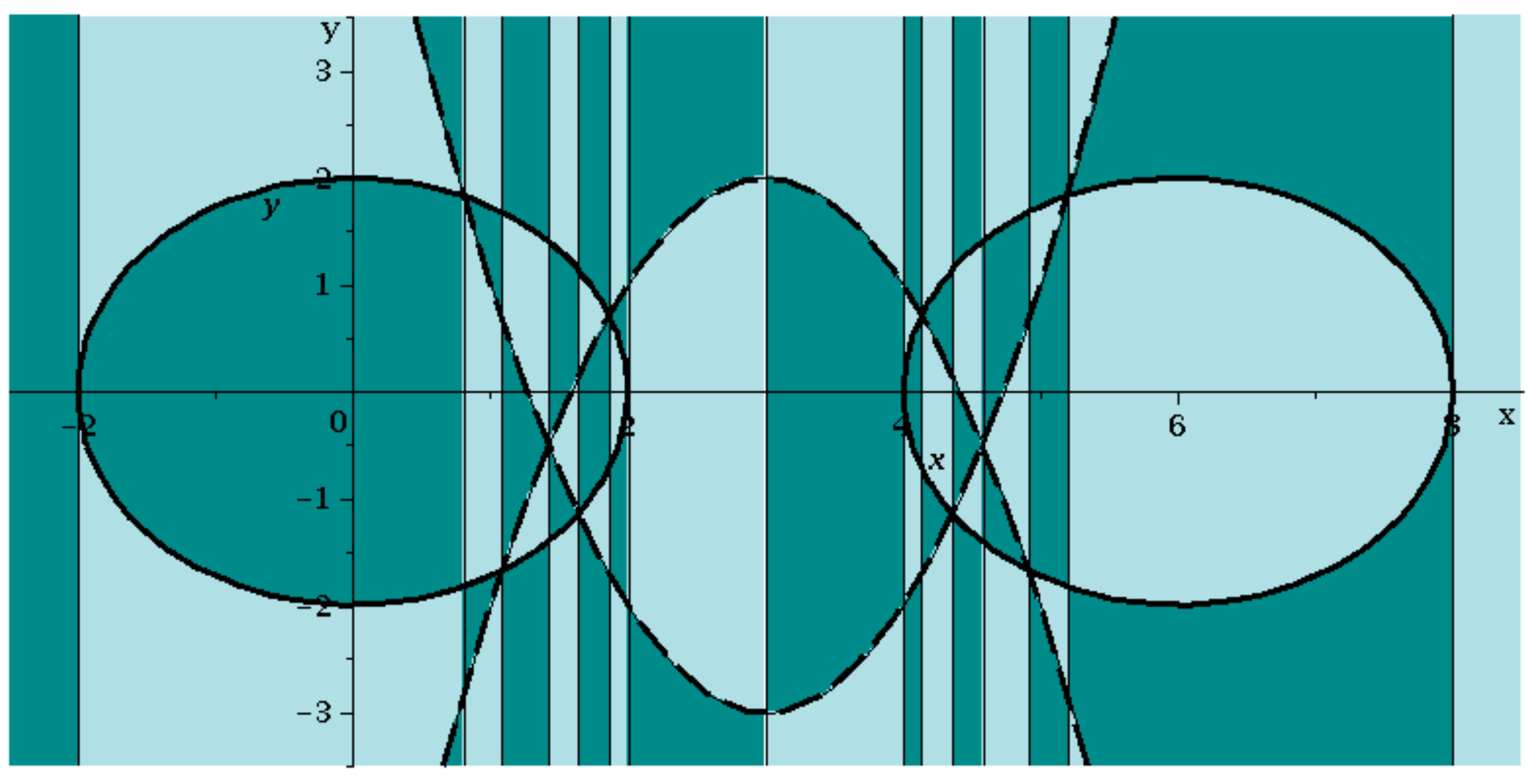}
\hspace*{0.02\textwidth}
\includegraphics[width=0.48\textwidth]{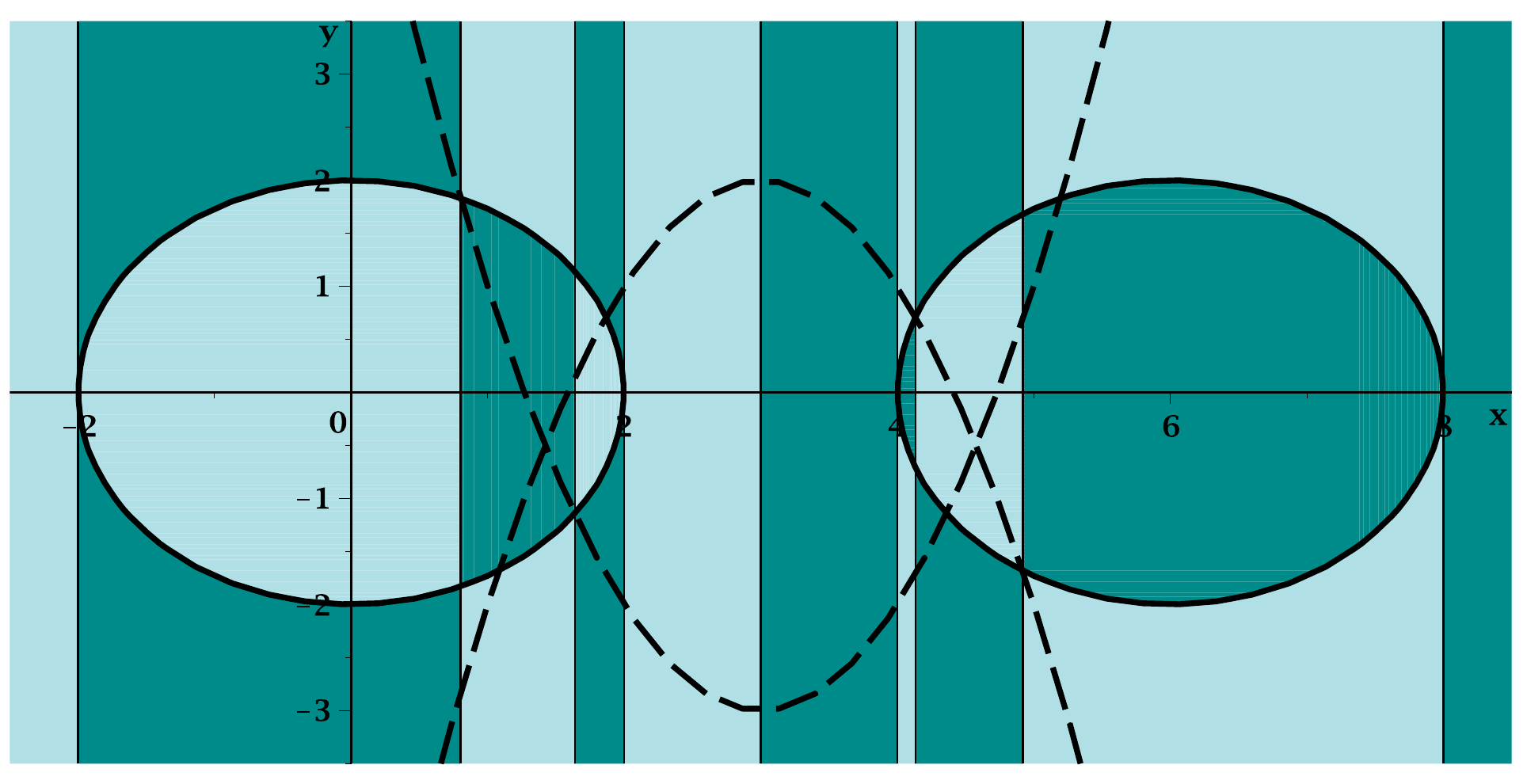}
\end{figure}

\noindent \textbf{Example:}  Consider (\ref{eq:TTI}) with 
\[
g_1 := x^2+y^2-4, \qquad g_2 := (x-3)^2-(y+3),
\]
and
\[
g_3 := (x-3)^2+(y-2),  \qquad g_4 := (x-6)^2+y^2-4.
\]
Figure \ref{fig:Ex1} shows the two dimensional cells produced for both a sign-invariant CAD and a truth-table invariant CAD, built under ordering $x \prec y$.
The sign-invariant CAD has 231 cells (72 full-dimensional but the splitting of the final cylinder is out of view) and the TTICAD 67 (22 full-dimensional).

By comparing the figures we see two types of differences.  First, the CAD of the real line is split into fewer cells (there are not as many cylinders in $\R^2$).  This is the effect of the reduction in projection polynomials identified, (less univariate polynomials with real roots to isolate).  The second difference is that the full-dimensional cylinders are no longer split over the dashed lines. This came from an improvement in the lifting phase (discussed in detail in \cite{Bradfordetal2014a}).  It leverages the projection theory to conclude that we usually only need to lift with respect to equational constraints themselves.

\vspace*{0.1in}

More recently \cite{Bradfordetal2014a} truth-table invariance has been achieved even when there is no implicit equational constraint, as with an example of the form 
\begin{equation}\label{eq:TTI2}
\left((h_1=0)\land (h_2>0)\right)\lor(h_3>0).
\end{equation}
The savings that can be achieved depend on the number of equational constraints involved in sub-clauses of the parent formula.
\par
It is also possible to apply equational constraints in the regular chains technology view of CAD \cite{ChenMorenoMaza2012a}, again even when there is no global equational constraint, as in (\ref{eq:TTI2}) \cite{Bradfordetal2014b}. 

\section{How Reliable is this?}

Cylindrical algebraic decomposition can be used as tool in program verification, as in the MetiTarski tool \cite{Paulson2012b}. This leads to the question: who will verify the CAD, or at least the inferences we draw from it?  We note that a positive answer to a purely existential question (equally, a negative answer to a purely universal question) is easily verified since we have a witness. The converse questions are essentially questions of refutation, see \cite{JovanovicdeMoura2012a}. Questions involving a mixture of quantifiers are much harder.

Almost all current implementations of CAD are based on computer algebra systems, which are generally unverified. We can at least compare, on a fairly level playing field, the implementations in \textsc{Maple} of four algorithms: see Table \ref{tab:compare}. The classification of the amount of mathematics involved is subjective, but we note that \cite{McCallum1984}, and hence \cite{Bradfordetal2013a}, relies on \cite{Zariski1965,Zariski1975} to justify the smaller projection set compared with \cite{Collins1975}. \cite{ChenMorenoMaza2012a} and  \cite{Bradfordetal2014b} rely on, {\it inter alia\/}, \cite{Aubryetal1999}.
\begin{table}
\caption{Comparison of algorithms\label{tab:compare}}
\centering
\begin{tabular}{llrr}
Algorithm&Implementation&Code Lines&Specialist\cr
&&(above Maple)&Mathematics\cr
\cite{Collins1975}&\cite{EWBD14}&
2600&some\cr
\cite{McCallum1984}&\cite{EWBD14}&
2500&a lot\cr
\cite{ChenMorenoMaza2012a}&\cite{ChenMorenoMaza2012a}&5000&medium\cr
\cite{Bradfordetal2013a}&\cite{EWBD14}&
3000&a lot\cr
 \cite{Bradfordetal2014b}& \cite{Bradfordetal2014b}&5500&medium\cr
\end{tabular}
\end{table}
\par
There are two challenges involved in verifying a CAD algorithm.
\begin{enumerate}
\item There is a ``program verification'' question of ensuring that the algorithms produce the result that they say they will, i.e. that resultants, discriminants, real roots etc. are computed correctly.  This is non-trivial, to say the least, sitting on top of an unverified computer algebra system, but should be feasible for an implementation based on a sound kernel, such as Coq or Isabelle.
\item There is a ``mathematics verification'' question whether the resulting decomposition is truly sign-/order-/truth table-invariant for the inputs. This is where the column labelled ``Mathematics'' in Table \ref{tab:compare} comes in.  The only  attempt to produce verified CADs known to the authors, \cite[in Coq]{CohenMahboubi2012a}, is based, not on \cite{Collins1975} and its successors, but rather on \cite[chapter 2]{Basuetal2006}, itself essentially that of \cite{Tarski1951}.
\item[2a.]There is an interesting tension here between ``precomputed'' and {\it ad hoc\/} verification. An implementation based in \cite{McCallum1984} would essentially have to verify the relevant theorems from \cite{Zariski1965,Zariski1975}, but these could be imported as pre-verified lemmas. An implementation based on \cite{ChenMorenoMaza2012a} would verify that \emph{in this case} we had an appropriate cylindrical decomposition of $\C^n$ which \emph{in this case} translated to an appropriate cylindrical algebraic decomposition of $\R^n$. 
\end{enumerate}

\section{Final thoughts}

The topics we focussed on in this paper are implemented in \textsc{Maple}: 
\begin{itemize}
\item CAD by Regular Chains is implemented in the \textsc{RegularChains} Library.  A version of this ships with the core \textsc{Maple} distribution while the latest version is freely available from \texttt{http://www.regularchains.org/}.  
\item The authors' own work (equational constraints, truth-table invariance, sub-decompositions) is freely available in a \textsc{Maple} package \textsc{ProjectionCAD}.  The latest version is available from: \texttt{http://opus.bath.ac.uk/43911/}.
\end{itemize}  

\noindent Other implementations of cylindrical algebraic decomposition include:
\begin{itemize}
\item \textsc{Mathematica} \cite{Strzebonski2006}; The commands \texttt{CylindricalDecomposition} and \texttt{Reduce} can make use of an underlying CAD implementation.  These commands can be exceptionally fast but it can be hard to judge the CAD components individually as they are just one of several underlying methods available and the output is in the form of formulae rather than cells.  
\item \textsc{Qepcad} \cite{Brown2003}; a dedicated interactive command-line program available from \texttt{http://www.usna.edu/CS/qepcadweb/B/QEPCAD.html}.  One notable feature is the SLFQ program which can simplify large quantifier free formulae giving more readable output.  \textsc{Sage} now has a \textsc{Qepcad} interface.
\item \textsc{Redlog} \cite{SS03}; this \textsc{Reduce} package implements CAD along with other quantifier elimination methods such as virtual substitution. 
\item \textsc{SyNRAC} \cite{IYAY13}; a \textsc{Maple} package notable for its symbolic-numeric approach.  An older version is available for free download from: \\ 
\texttt{http://jp.fujitsu.com/group/labs/en/techinfo/freeware/synrac/} \\
with more recent advances part of the wider Todai Robot project.
\end{itemize}

\noindent The only reported experiments to cover all of these implementations were detailed in Section 4 of \cite{Bradfordetal2014b}.

Of course, this paper surveyed only a few of the recent advances in cylindrical algebraic decomposition.  Others include (but are not limited to):
\begin{itemize}
\item The use of certified numerics in the lifting phase to minimise the amount of symbolic computation required \cite{Strzebonski2006,IYAY13}.
\item Local projection schemes \cite{Strzebonski2014a}, generic projection schemes \cite{SS03} and single CAD cells \cite{Brown2013,JovanovicdeMoura2012a}.
\item Problem formulation for CAD \cite{Dolzmannetal2004a,Bradfordetal2013a,WEBD14} (projection and lifting) \cite{Englandetal2014c,EBCDMW14} (regular chains).  These all develop heuristics to help with choices, while \cite{Huangetal2014a} applies machine learning in the form of support vector machines to pick a heuristic.
\item Work on cylindrical algebraic sub-decompositions, which return only a subset of the cells in a full CAD \cite{Sei06}.  In \cite{WBDE14} algorithms are given to return cells that lie on a prescribed variety, or have a designated dimension, while in \cite{WDEB13} these techniques are combined to solve a motion planning problem.  Note that if restricting to cells of full dimension then sample points can always be chosen to be rational, greatly reducing running time.
\end{itemize}

There are numerous unsolved problems, both theoretical and practical.  Three that stand out to the authors are the following.
\begin{enumerate}
\item There is no complexity analysis of the Regular Chains method (though clearly it is subject to the lower bounds in Section \ref{sec:LB}).
\item There has been much progress in the last forty years, but implementations (at least for systems with alternations of quantifiers) are still doubly-exponential in the number of variables while the theory suggests we can do better.
\item Cylindricity is needed in step \ref{step:bool} of quantifier elimination, as $\exists$ translates into $\bigvee$ and $\forall$ into $\bigwedge$. However, in fact we only need this at the points where $\exists$ and $\forall$ alternate, so we can weaken the definition of cylindricity from being true for all $\pi_k$ to merely being true for those $k$ where $x_k$ and $x_{k+1}$ are governed by different quantifiers (or where $x_k$ is unquantified but $x_{k+1}$ is quantified, a concept we can call \emph{block-cylindrical}. Unfortunately, we currently know of no way of computing a block-cylindrical algebraic decomposition without computing the full cylindrical algebraic decomposition first.
\end{enumerate}

\subsection*{Acknowledgements}

This work was supported by the EPSRC (grant number EP/J003247/1).  

The authors thank Russell Bradford,
Nicolai Vorobjov, David Wilson (University of Bath), Changbo Chen (Chinese Academy of Sciences, Chongqing), Zongyan Huang (University of Cambridge), Scott McCallum (Macquarie University) and Marc Moreno Maza (Western University).
\bibliography{../../../jhd}

\end{document}